\documentclass[
  ,final            
  ]
  {aipproc}

\layoutstyle{8x11single}

\begin{document}

\def\Swift{\emph{Swift}}
\newcommand{\lesssim}{\lower.5ex\hbox{$\; \buildrel < \over\sim \;$}}
\newcommand{\gtrsim}{\lower.5ex\hbox{$\; \buildrel > \over\sim \;$}}
\newcommand{\apj}{{\it Astrophysical J.\ }}
\newcommand{\apjs}{{\it Astrophysical J.\ Supp.\ }}
\newcommand{\apjl}{{\it Astrophysical J.\ Lett.\ }}
\newcommand{\mnras}{{\it MNRAS~}}
\newcommand{\aap}{{\it Astron.\ Astrophys.\ }}
\newcommand{\aaps}{{\it Astron.\ Astrophys.\ Supp.\ }}

\title{The Extragalactic $\gamma$ Ray Background}


\author{Charles D. Dermer}{
  address={Space Science Division, Code 7653,
Naval Research Laboratory\break Washington, DC 20375-5352, USA} }

\begin{abstract}
One way to understand the nonthermal history of the universe is by
establishing the origins of the unresolved and truly diffuse
extragalactic $\gamma$ rays. Dim blazars and radio/$\gamma$ galaxies
certainly make an important contribution to the galactic $\gamma$-ray
background given the EGRET discoveries, and previous treatments are
reviewed and compared with a new analysis.  Studies of the
$\gamma$-ray intensity from cosmic rays in star-forming galaxies and
from structure formation shocks, as well as from dim GRBs, are briefly
reviewed. A new hard $\gamma$-ray source class seems required from the
predicted aggregate intensity compared with the measured intensity.
\end{abstract}

\maketitle


\section{Introduction}

An isotropic, apparently diffuse flux of $\gamma$ rays was discovered
with SAS-2 in the $\approx 40$ -- 200 MeV range \citep{tf82}. EGRET,
improving and extending the SAS-2 result, measured isotropic
$\gamma$-ray emission in the $\approx 30$ MeV -- 100 GeV range
\citep{sre98} with $\nu F_\nu$ intensity at 1 GeV at the level 
of $\approx 1$ keV/(cm$^{2}$-s-sr), and with $\nu F_\nu$ spectral
index $\alpha_\nu \approx -0.10\pm0.03$ (Fig.\ 1a).  The diffuse
isotropic $\gamma$-ray background consists of an extragalactic
$\gamma$-ray background and an uncertain contribution of
quasi-isotropic Galactic $\gamma$ rays produced, for example, by
Compton-scattered radiations from cosmic-ray electrons.  The
model-dependent Galactic contribution
\citep{sre98,smr00,smr04,smr04a}, and the addition at some level of 
heliospheric flux \citep{mpd06,ops07}, means that the actual
contribution from extragalactic sources is somewhat uncertain. The
data in Fig.\ 1a compares the extragalactic diffuse $\gamma$-ray
intensity from EGRET analysis \citep{sre98} with results using the
GALPROP model \citep{smr04a}, the latter of which requires an extended
($\approx 4$ -- 10 kpc) nonthermal electron halo to fit the hard
($\alpha_\nu \cong -0.4$) diffuse Galactic $\gamma$-ray emission.  For
our purposes, we consider the apparently diffuse extragalactic
$\gamma$-ray background (EGRB) of \citet{smr04a} as the conservative
upper limit for the superposed intensity of any class of $\gamma$-ray
sources, with the \citet{sre98} intensity as an absolute upper limit
to the combined residual intensity from all source classes.

The GALPROP fits \cite{smr00} to the OSSE-COMPTEL-EGRET Milky Way
intensity spectra in different directions toward the Galaxy implies
the total $\gamma$-ray luminosity of the Milky Way galaxy. Scaled to
$10^{39}L_{39}$ ergs s$^{-1}$, the GALPROP analysis gives $L_{39} = (0.71 -
0.92)$ for the $> 100$ MeV $\gamma$-ray luminosity of the Milky Way,
a factor $\approx 3$  greater than the value $L_{39} =(0.16 - 0.32)$
inferred from COS-B observations \citep{blo84}. Most of this emission
is from secondaries created in cosmic-ray nuclear production
processes.  The Galactic $\gamma$-ray power provides an important
yardstick to assess the total contribution of to the unresolved
$\gamma$-ray background of cosmic-ray emissions from star-forming
galaxies, as described in more detail below.

Every $\gamma$-ray source class makes a different contribution to the
$\gamma$-ray background, including transient events below detector
threshold, variously oriented relativistic jet sources, and large
numbers of individually weak sources.  The basic formalism for making
such calculations for beamed and unbeamed sources was given in my
Barcelona talk \citep{der06}. Here I review the various source
classes that likely dominate the composition of the diffuse
background: blazars and radio/$\gamma$ galaxies; star-forming galaxies
of various types; $\gamma$ rays from structure-formation shocks; and
GRBs.

\section{Blazars and Radio/$\gamma$ Galaxies}

Population studies of $\gamma$-ray blazars were undertaken soon after
the recognition of the $\gamma$-ray blazar class with EGRET \citep{fic94}. 
\citet{chi95} performed a $\langle V/V_{max}\rangle$ analysis assuming
no density evolution and showed that luminosity evolution of EGRET
blazars was implied by the data. With a larger data set, and using
radio data to ensure the sample was unbiased in regard to redshift
determination, \citet{cm98} again found that luminosity evolution was
required. They obtained best-fit values through the maximum likelihood
method that gave an AGN contribution to the EGRET $\gamma$-ray
background at the level of $\approx 25$\%.

\begin{figure}[t]
\includegraphics[width=6.4in]{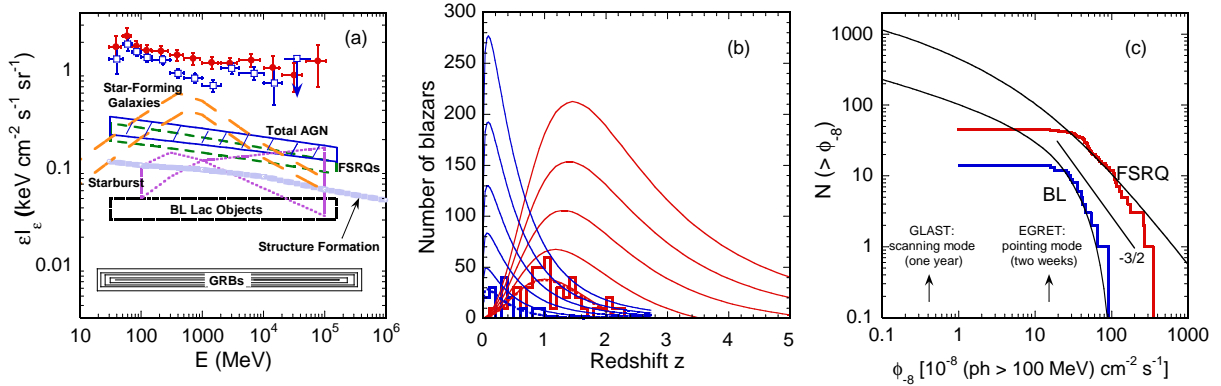}
\caption{{\small (a) Diffuse extragalactic $\gamma$-ray background from 
analyses of EGRET data, shown by filled \cite{sre98} and open
\cite{smr00} data points, compared to model calculations of the
contributions to the EGRB for FSRQs and BL Lac objects, and total AGNs
\cite{der06}, star-forming galaxies \cite{pf02}, starburst galaxies
\cite{tqw07}, structure shocks in clusters of galaxies
\cite{kes03,bgb07}, and GRBs \cite{der07}. (b) Fitted EGRET and
predicted redshift distributions of FSRQs and BL Lac objects
\citep{der07}. (c) Fitted EGRET size distribution, and predictions for
different flux levels \citep{der07}.}}
\label{fig3} 
\end{figure}

\citet{ss96} postulated a radio/$\gamma$-ray correlation in blazars, and
tried to correct for the duty cycle and $\gamma$-ray spectral
hardening of flaring states.  They found that essentially 100\% of the
EGRET $\gamma$-ray background arises from unresolved blazars and AGNs.
In later work \citep{ss01}, they predict that GLAST will detect
$\approx 5000$ blazars to a flux level of $\approx 2\times 10^{-9}$
ph($> 100$ MeV)/(cm$^2$-s), which will be reached with GLAST after
$\approx 4$ years.  They did not, however, fit the blazar redshift
distribution to provide a check on their model, nor distinguish
between flat spectrum radio quasar (FSRQ) and BL Lac objects.

The crucial underlying assumption of this approach, which has been
developed in recent work \citep{gio06,nt06}, is that there is a simple
relation between the radio and $\gamma$-ray fluxes of blazars. Because
a large number of EGRET $\gamma$-ray blazars (primarily FSRQs) are
found in the 5 GHz, $>1$ Jy \citet{kue81} catalog, a
radio$/\gamma$-ray correlation is expected. This correlation is not,
however, evident in 2.7 and 5 GHz monitoring of EGRET $\gamma$-ray
blazars \citep{mue97}. X-ray selected BL objects are also not
well-sampled in GHz radio surveys. Studies based on correlations
between the radio and $\gamma$-ray emissions from blazars must
therefore consider the very different properties and histories of
FSRQs and BLs and their separate contributions to the $\gamma$-ray
background.

Treatments of blazar statistics that avoid any radio/$\gamma$-ray
correlation and separately consider FSRQs and BL Lac objects have been
developed by \citet{mp00} and \citet{der07}. In the \citet{mp00}
study, blazar spectra were calculated assuming an injection electron
number index of $-2$.  Distributions in injected particle energy in BL
Lac and FSRQ jets were separately considered, with a simple
description of density evolution given in the form of a cutoff at some
maximum redshift $z_{max}$. Depending on the value of $z_{max}$,
\citet{mp00} concluded that as much as $\approx 40$
-- 80\% of the EGRB is produced by unresolved AGNs, with 
$\approx 70$ -- 90\% of the emission from FR 1 galaxies and
BL Lac objects.

In my  recent study \citep{der07}, I also  use a physical model to fit
the EGRET  data  on  the  redshift  and  size  distribution  of  EGRET
blazars. The EGRET   blazar sample consists of 46 FSRQs and 14 BL  Lac
objects that were detected in the Phase 1 EGRET all-sky survey
\citep{fic94}, with fluxes as reported in the Third EGRET catalog
\citep{har99}. A blazar is approximated by a relativistic spherical
ball entraining a tangled magnetic field and containing an isotropic,
power-law distribution of nonthermal electrons. Single electron
power-law distributions were used in the study, with indices $p = 3.4$
for FSRQs and $p=3.0$ for BL Lac objects, giving spectral indices
$\alpha_\nu = -0.2$ and $\alpha_\nu = 0.0$, respectively, as shown by
observations \citep{muk97,vp07}. Beaming patterns appropriate to
external Compton and synchrotron self-Compton processes, and bulk
Lorentz factor $\Gamma = 10$ and $\Gamma = 4$, were used in FSRQs and
BL Lac objects, respectively.  The comoving directional luminosities
$l_e^\prime$ and blazar comoving rate densities (blazar formation
rate; BFRs) for the two classes were adjusted to give agreement with
the data.  The threshold detector sensitivity $\phi_{-8}$, in units of
$10^{-8}$ ph$(> 100$ MeV)/(cm$^2$-s), was nominally taken to be
$\phi_{-8} = 15$ for the two-week on-axis EGRET sensitivity, and
$\phi_{-8} = 0.4$ for the one-year all-sky sensitivity of GLAST.  Due
to incompleteness of the sample near threshold, the EGRET threshold
was adjusted to $\phi_{-8} = 25$.  Because a mono-luminosity function
was used, the range in apparent powers is entirely kinematic in this
model, arising from the different, randomly oriented jet directions.

By using a minimalist blazar model, the model parameters were severely
constrained. The FSRQ data were fit with $l_e^\prime = 10^{40}$
ergs/(s-sr) and a BFR that was $\approx 15\times$ greater at $z
\approx 2$ -- 3 than at present. The BL Lac data, by contrast, could
not be fit using a fixed luminosity.  A model that could jointly fit
the redshift and size distribution of BL Lac objects required that BL
Lac objects be brighter and less numerous that in the past, consistent
with a picture where FSRQs evolve into BL Lac objects \citep{bd02,mgt07}.

Fig.\ 1b shows the fitted EGRET redshift distributions and predicted
redshift distributions of $\gamma$ galaxies and blazars at different
GLAST sensitivities \citep{der07}.  The fits to the EGRET size
distributions of FSRQs and BL Lac objects, and extrapolations of the
model size distributions to lower flux thresholds, are shown in Fig.\
1c.  After one year of observations with GLAST ($\phi_{-8} \cong
0.4$), $\approx 800$ FSRQs/FR2 and $\approx 200$ BL Lac/FR1 $\gamma$
galaxies and $\gamma$-ray blazars are predicted. This is a lower
prediction, and additional hard-spectrum blazars to which EGRET was
not sensitive could increase this number, but not by more than a
factor $\approx 2$. The contribution of unresolved blazars below a
flux level of $\phi_{-8} \cong 12.5$ -- 25 to the EGRB is shown in
Fig.\ 1a. As can be seen, the total blazar/$\gamma$ galaxy
contribution is less than $\approx 20$ -- 30\% of the EGRET EGRB
intensity, meaning that other classes of sources must make a
significant contribution.

\section{Star-Forming Galaxies}

The integrated emission from $\gamma$ rays formed by cosmic-ray
interactions in star-forming galaxies will make a ``guaranteed''
$\gamma$-ray background. Pavlidou \& Fields \cite{pf02} calculate this
intensity by approximating the diffuse Galactic $\gamma$-ray spectrum
as a broken power law and assuming that the $\gamma$-ray spectrum of a
star-forming galaxy is proportional to the supernova rate and thus the
massive star-formation rate, which can be inferred from the measured
blue and UV luminosity density. Fig.\ 1a shows their results for a
dust-corrected star formation rate (SFR) integrated over all
redshifts, and a lower curve where the SFR is integrated to redshift
unity.

A different approach \citep{tqw07,tqwl06} to this problem starts by
noting that cosmic-ray protons in the Milky Way lose only $\approx
10$\% of their energy before escaping. This fraction could rise to
nearly 100\% in starburst galaxies where the target gas density is
much higher and the timescale for escape, due primarily to advective
galactic winds rather than diffusion in the galaxy's magnetic field,
is less than the nuclear loss time. Support for this contention is
provided by the observed correlation between far infrared
flux---primarily due to starlight reradiated by dust and gas---with
synchrotron flux produced by cosmic ray electrons. If both are
proportional to the supernova rate, and the radio-emitting electrons
lose a large fraction of their energy due to synchrotron cooling, then
this correlation is explained \citep{vol89}.

The calculated intensity \citep{tqw07} from starburst galaxies is
shown in Fig.\ 1a. The bulk of this intensity is formed at redshifts
$\gtrsim 1$, where the starburst fraction of star-forming galaxies is
large. The starburst intensity from Ref.\ \citep{tqw07} is smaller
than the the total star-forming galaxy contribution \citep{pf02}, even
when the latter calculation is truncated at $z = 1$. The latter
calculation was checked in Ref.\ \citep{der06}, based on the
$\gamma$-ray spectrum of the Milky Way. Stecker \citep{ste07} argues
that the starburst contribution is a factor $\approx 5$ lower than
diffuse $\gamma$-ray and neutrino intensity derived by Loeb and Waxman
\citep{lw06} and Thompson et al.\ \cite{tqw07} by pointing out that
directly accelerated electrons make a strong contribution to the
synchrotron flux, and questioning the assumption that protons lose all
their energy in starbursts. This criticism is addressed in 
Ref.\ \cite{tqwl06}.

GLAST will clarify this situation through its observations of nearby
star-forming galaxies, e.g., LMC, SMC, M31, and M33, the starburst
galaxies M82 and NGC 253, and infrared luminous galaxies like Arp
220. These galaxies are predicted to be GLAST sources
\citep{pf02,tor04,dt05,tqw07}, and will provide benchmarks to correlate
$\gamma$-ray fluxes with star formation activity.

\section{Clusters of Galaxies}

Nonthermal radiation from clusters of galaxies is expected for several
reasons: cosmic rays will be accelerated through merger shocks from
merging subclusters, from accretion shocks as primordial matter
continues to accrete on a forming cluster, and from turbulent
reacceleration of nonthermal particles by plasma waves in the
intracluster medium. In addition, a galaxy cluster often has an
energetic AGN in its central cD galaxy that could inject cosmic rays
into the cluster medium. Hadronic cosmic rays with energies $\lesssim
10^{19}$ eV will be trapped on timescales longer than the Hubble
time, so galaxy clusters become storage volumes for cosmic rays
\citep{bbp98}. In spite of these expectations, EGRET did not make a
high-significance detection of any galaxy cluster \citep{rei03}.

Hard X-ray tails have also not been detected with high significance
from the Coma cluster or any other galaxy cluster. The study of
nonthermal emission from clusters of galaxies has consequently
stalled, as nonthermal X-ray measurements provide the crucial
information to normalize the magnetic field and nonthermal electron
spectrum.  Predictions based on the marginal detection of the hard
X-ray tail from the Coma cluster indicate that Coma will be easily
detectable with GLAST in one year of observation and marginally
detectable with ground-based $\gamma$-ray telescopes in a nominal 50
hour observation \citep{bd04}, though the angular extent of Coma makes
such detections more difficult \citep{gb04}.

In view of these uncertainties, any calculation of the integrated
contribution from clusters of galaxies to the $\gamma$-ray background
is likewise highly uncertain. Fig.\ 1a shows predictions
\citep{kes03,bgb07} for galaxy cluster emission. GLAST detections 
of clusters of galaxies will be crucial to provide a better basis for
determining this contribution.

\section{Gamma Ray Bursts}

The contribution of untriggered GRBs to the $\gamma$-ray background
can be estimated in a number of ways, but all depend on modeling, or
inferring from observations, the typical high-energy GRB spectra. For
the optimistic case that the TeV flux made by a GRB is $\approx
10\times$ greater than the MeV flux, then the superpositions of GRB
emissions are found to make $\approx 10$\% of the $\gamma$-ray
background after cascading from high energies into the GeV band
\citep{cdz07}.  If one instead relies on observations of EGRET
spark-chamber GRBs that show that the fluence in the EGRET band is
only $\approx 10$\% of the fluence in the BATSE band, then GRBs are
found to give very little ($\lesssim 1$\%) contribution to the
$\gamma$-ray background \citep{ld07}. This neglects the contributions
of short, hard GRBs and low luminosity GRBs, but since these have
small fluences and all-sky rates, they are unlikely to make
a significant contribution to the $\gamma$-ray background.

It hardly needs to be mentioned that GLAST observations of the
high-energy emission from GRBs will provide crucial information to
determine the share of the background $\gamma$-ray intensity provided
by GRBs.

\section{Additional Contributions}

A truly diffuse flux of $\gamma$ rays will be formed by the cascade
radiations initiated by photopion and photopair production of
ultra-high energy cosmic rays interacting with photons of the
extragalactic background light. Because the electromagnetic
secondaries are distributed over several orders or magnitude as they
cascade to photon energies where the universe becomes transparent to
$\gamma\gamma$ processes, this intensity will be well below the
Waxman-Bahcall intensity at $\approx 0.03$ keV/(cm$^2$-s-sr).  By
comparing with the diffuse neutrino intensities calculated in
bottom-up scenarios for the ultra-high energy cosmic rays
\citep{der07a,yk06}, cosmogenic $\gamma$ rays are not expected to make
a large contribution to the EGRB (however, see Ref.\ \citep{kss07},
though they could for top-down models \citep{ss04}. This question will
definitively be answered by Auger data.

The various source classes that contribute to the extragalactic
$\gamma$-ray background have hardly been exhausted, but the described
classes are expected to be most important. Yet when one adds up the
best guesses of the various contributions to the total, as shown in
Fig.\ 1a, a deficit remains at both low ($\lesssim 100$ MeV) and high
($\gg 1$ GeV) energies. Because star-forming and starburst galaxies
make such a large contribution to the total, it is possible that their
spectra are actually much softer than assumed on the low-energy side,
due to nonthermal electron bremsstrahlung and Compton-scattered
emissions from $\gamma$-ray production by cosmic rays in
``thick-target' starburst and infrared luminous galaxies (cf.\
\citep{mdr87}). This, or soft-spectrum radio galaxies and 
from the superposition of hard tailes from many weak radio-quiet
Seyfert galaxies, could explain the low-energy deficit.

It seems unlikely, however, that star-forming galaxies, whose
high-energy radiation originates from cosmic rays accelerated by
supernova remnant shocks, could explain the deficit on the high-energy
side unless shock injection spectra harder than $-2$ were
postulated. The EGRET effective area dropped rapidly above $\approx 5$
GeV due to self-vetoing effects, so it was not sensitive to
hard-spectrum sources, in particular, hard spectrum BL Lac
objects. But the BL Lac contribution is estimated at the 5\% level,
and it is difficult to suppose that EGRET was not able to detect a
number of such hard-spectrum BL Lac objects. Hard tails on FSRQs
originating, e.g., from photohadronic cascade emissions \citep{ad03},
could explain the high-energy discrepancy. Other possibilities are the
diffuse contributions from dark matter annihilation \citep{ull02}, or
cascade radiations from misaligned blazars \citep{acv94}. We must
furthermore keep in mind the possibility that the model of foreground
Galactic emission that must be subtracted from the extragalactic flux
is incomplete \citep{kes04}, or that the EGRET internal background was
underestimated \citep{atw07}. Data from GLAST will tell us which, if
any, of these suggestions are correct, and whether new, unexpected
sources of high-energy $\gamma$ rays are required to explain the
$\gamma$-ray background.


\begin{theacknowledgments}
The work of C.~D.~D. is supported by the Office
of Naval Research and NASA {\it GLAST} Science Investigation
No.~DPR-S-1563-Y.
\end{theacknowledgments}

\end{document}